\documentclass[aps,prb,showpacs,twocolumn,preprintnumbers,superscriptaddress,amsmath,amssymb,amsfonts,floatfix]{revtex4-1}

\usepackage{graphicx, color,rotating, textcomp}
\usepackage{dcolumn}
\usepackage{bm}

\begin{document}

\title{Quantum Hall effect in graphene with twisted bilayer stripe defects}

\author{Tomas~L\"ofwander}
\affiliation{Department of Microtechnology and Nanoscience - MC2,
Chalmers University of Technology, SE-412 96 G\"oteborg, Sweden}

\author{Pablo San-Jose}
\affiliation{Instituto de Estructura de la Materia (IEM-CSIC),
Serrano 123, 28006 Madrid, Spain}

\author{Elsa Prada}
\affiliation{Instituto de Ciencia de Materiales de Madrid, CSIC,
Cantoblanco, 28049 Madrid, Spain}

\date{\today}

\begin{abstract}
  We analyze the quantum Hall effect in single layer graphene with
  bilayer stripe defects.  Such defects are often encountered at steps
  in the substrate of graphene grown on silicon carbide.  We show that
  AB or AA stacked bilayer stripes result in large Hall conductivity
  fluctuations that destroy the quantum Hall plateaux.  The
  fluctuations are a result of the coupling of edge states at opposite
  edges through currents traversing the stripe.  Upon rotation of the
  second layer with respect to the continuous monolayer (a
  twisted-bilayer stripe defect), such currents decouple from the
  extended edge states and develop into long-lived discrete quasi
  bound states circulating around the perimeter of the stripe.
  Backscattering of edge modes then occurs only at precise resonant
  energies, and hence the quantum Hall plateaux are recovered as twist
  angle grows.
\end{abstract}

\pacs{
73.50.Jt %Hall effect in Thin Films
72.80.Vp %Electronic transport - graphene
85.75.Nn %Hybrid Hall effect devices
}

\maketitle

\section{Introduction}

The unique half-integer quantum Hall effect (QHE) in monolayer
graphene serves as a fingerprint of massless Dirac
electrons. \cite{Novoselov:N05,Zhang:N05} It is therefore used in the
laboratory to distinguish monolayers from
multilayers. \cite{Novoselov:NP06} The electrons in graphene under
applied perpendicular magnetic field have an unconventional Landau
level spectrum, leading to a sequence of Hall conductivity plateaux
$\sigma_{xy}=G_0 (2n+1)$, where $G_0$ is the conductance quantum,
$G_0=2e^2/h$ (h is Planck's constant and e is the electron charge),
and $n$ is an integer including zero. \cite{Gusynin:PRL05} The large
energy level separation between the $n=0$ and $n=1$ Landau levels adds
robustness to the $n=0$ plateau, which has been observed also at room
temperature. \cite{Novoselov:S07} More importantly, measurements
\cite{Tzal:Nature10,Poirier:CRP11,Janssen:M12} of the von Klitzing
constant $R_K=h/e^2$ have been performed to metrological accuracy on
epitaxial graphene on silicon-carbide (SiC). Large break-down currents
have been observed for this material, and epitaxial graphene at
present outperforms conventional two-dimensional electron gases in
semiconducting heterostructures in this respect, and may very well be
the material of choice for metrology in the future.  Transistors with
promising high cut-off frequencies have also been fabricated from
epitaxial graphene. \cite{Lin:S10,Liao:N10} It is therefore of high
current interest to establish the electron transport properties of
graphene on SiC. \cite{First:MB10}

\begin{figure}[b]
\includegraphics[width=0.8\columnwidth]{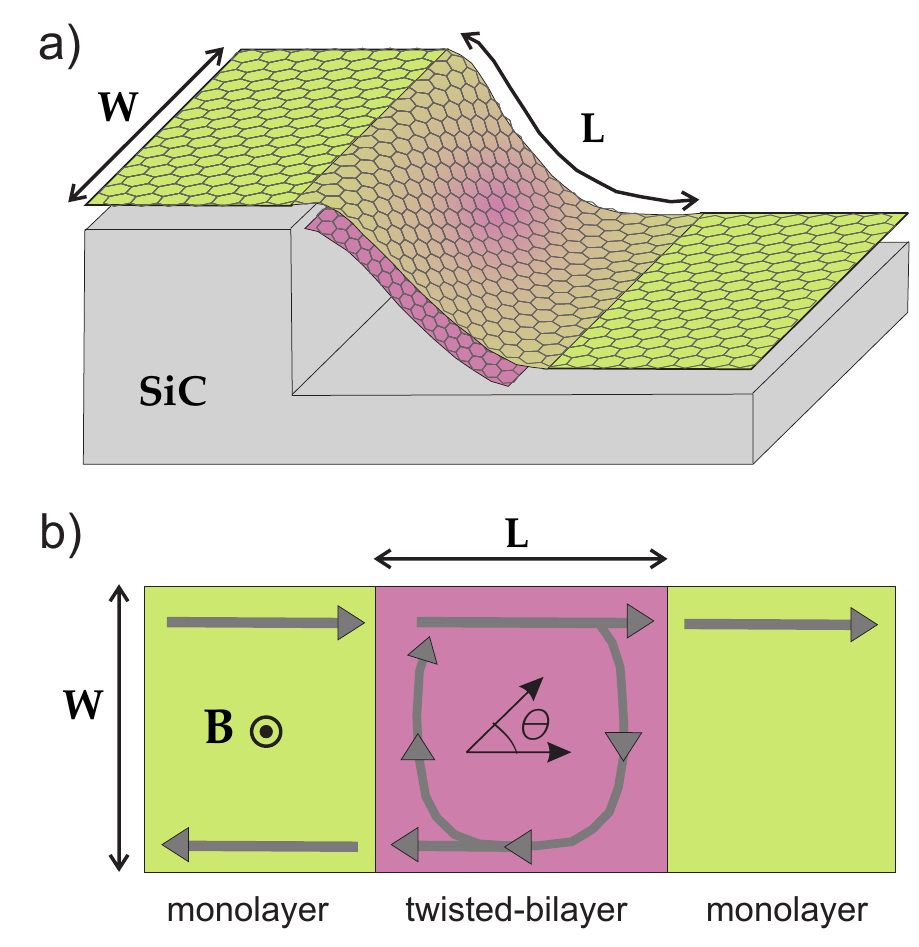}
\caption{(a) Illustration of a continuous graphene layer over a
  substrate with two terraces separated by a step.  A second layer is
  formed at the step. (b) Schematics of a graphene monolayer-twisted
  bilayer-monolayer junction in a perpendicular magnetic field. The
  ribbon's width is $W$, the bilayer patch has a length $L$ and the
  lattice twist angle between layers is $\theta$. Allowed edge state
  paths for electrons are sketched in each region.}
\label{fig_system}
\end{figure}

Inhomogeneities in the two-dimensional material are often detrimental
to its transport properties.
\cite{Tedesco:APL09,Robinson:NL09,Dimitrakopoulos:APL11,Bryan:JPCC11,Desh:PhysE12}
Epitaxial graphene on SiC may continuously cover the whole SiC
substrate, \cite{Seyller:SS06,Jobst:PRB10,Jobst:SSC11} but steps on
the substrate influence the graphene layer along lines running across
the wafer. \cite{Yakes:NL10,Ji:NM12} At a step, the graphene sheet may
be more decoupled from the underlaying substrate than on the wide
terraces between steps, which may change the doping level
locally. \cite{Yakes:NL10,Low:PRL12} Graphene may also suffer strain
\cite{Prada:PRB10} since the SiC step is atomically sharp, while the
graphene sheet forms a continuous cover.  In addition, since the steps
serve as seeds in the growth process of epitaxial graphene, bilayers
or multilayers are often observed,
\cite{Viro:PRB08,Emtsev:NM09,Jobst:SSC11} see Fig. \ref{fig_system}
(a). Depending on the growth process, several islands may form near
the steps or continuous stripes may be formed along a large part of
the step.  After fabrication of the Hall bar, the bilayer stripe
defects can reach from one side to the other of the Hall bar,
resulting in a geometry similar to the one shown in
Fig.~\ref{fig_system} (b).  Experimentally, it was recently observed
\cite{Schumann:PRB12} that narrow Hall bars intentionally fabricated
perpendicular or parallel to steps display markedly different
properties. When the current path crosses many steps, a positive
magnetoresistance arises, that can be explained, according to Schumann
et al., \cite{Schumann:PRB12} as the result of Hall edge channel
backscattering caused by new edge channels developing along the
substrate steps, although the specific mechanism remains an open
question. In contrast, other experiments \cite{Jobst:SSC11,
  Pan:APL10,Tanabe:APE10} show that the magnetotransport in epitaxial
graphene appears basically insensitive to surface steps. Thus, a
theory that embraces both scenarios is still missing.

Here, we numerically investigate a scenario that reconciles both
observations, wherein a bilayer patch interferes with the currents
flowing in the underlaying monolayer, as sketched in
Fig.~\ref{fig_system} (b). We show that the QHE plateau quantization
is strongly suppressed by the presence of a single AA- or AB-stacked
bilayer stripe crossing the Hall bar, which opens up the possibility
of edge state backscattering by connecting opposite edges. This
effect, however, becomes much weaker as the two layers are rotated by
a finite relative angle, breaking the perfect AA or AB stacking.  We
find that the QHE is least distorted as the twist angle approaches
$30^\circ$ (midway between AA and AB stacking). Although inter-edge
backscattering remains possible in this case, it becomes confined to
narrow resonances, apparent as narrow dips in the Hall plateaux, and
caused by quasi-bound states circulating around the patch that are
weakly coupled to the extended edge states. The backscattering
resonances are furthermore smeared out by finite temperature
effects. Hence, a significant suppression of Hall plateaux in
SiC-grown epitaxial graphene typically requires the Hall bar to lie
across substrate steps, as found in Ref. \onlinecite{Schumann:PRB12},
but also good crystallographic alignment of the multilayer patches
seeded by the steps.

\section{Hall effect across a twisted bilayer} 
The properties of bilayer graphene, particularly of twisted bilayers,
have been the focus of considerable interest
recently. \cite{Santos:PRL07,Bistritzer:P11,San-Jose:PRL12,Santos:PRB12}
For AB-stacked bilayer graphene, the two Dirac cones of a decoupled
double monolayer system are strongly modified by the interlayer
hopping, resulting in parabolic bands and possibly trigonal
warping. \cite{McCann:SSC07} In twisted graphene, on the other hand,
the two cones within each valley are separated in reciprocal space and
interlayer coupling leads to a finite energy saddle point in the band
structure at the intersection of the two surviving Dirac cones. The
corresponding van Hove singularity has been observed
experimentally. \cite{Li:NP10} The question arises as to what the QHE
looks like across a monolayer-bilayer graphene junction, including the
effect of interlayer twist in the bilayer part.

It should be recognized that a heterostructure\cite{Tsukuda:2011}
consisting of monolayer graphene occupying the half space $x<0$ and
bilayer graphene occupying the other half space $x>0$ is rather
different from the geometry considered in this paper, where the
bilayer exists between $0<x<L$, see Fig.~\ref{fig_system}, and plays
the role of a complicated barrier for electron flow in the lower
extended monolayer. The finite length $L$ of the bilayer patch leads
to the formation of a spectrum with quasi-bound state resonances. Such
states are chiral and circulate around the perimeter of the patch, but
may escape into the two extended states at opposite edges of the
monolayer (and eventually to reservoirs) through two opposite corners
of the patch [see Fig. \ref{fig_system} (b)]. When the Fermi energy
equals a resonance energy, a vertically propagating channel is opened
that connects an incoming edge state into an outgoing state at the
opposite edge, allowing for backscattering in the QHE regime. This
appears as a dip of depth $G_0$ in the quantized value of the Hall
conductivity across the resonance. If the width of the resonance
levels exceeds the corresponding level separation, the Hall
conductivity plateaux are completely destroyed.  Ultimately, the
existence of such transverse backscattering channels has a topological
origin, since the different Chern-numbers of the monolayer and bilayer
bands dictates that the number of edge channels along a
monolayer-bilayer interface is odd, as a consequence of the
bulk-surface correspondence. \cite{Hasan:RMP10,Prada:SSC11}

\begin{figure}[t]
\includegraphics[width=\columnwidth]{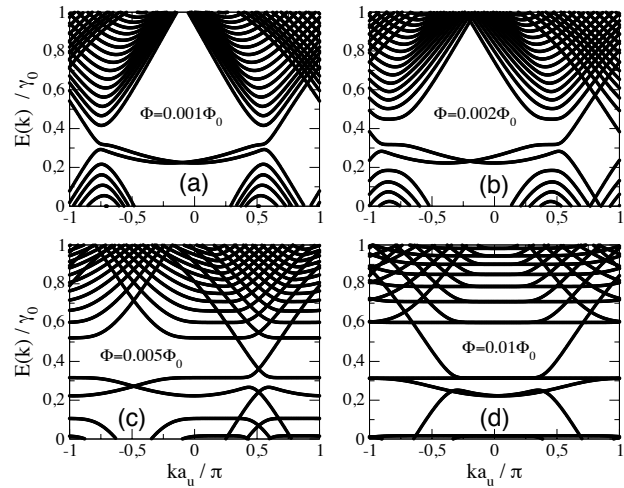}
\caption{Bandstructure of a 10 nm wide zigzag graphene ribbon at
  various magnetic fields for a hopping cut-off of $R_c=3a_{cc}$ and
  ribbon unit cell size $a_u=2\sqrt{3}a_{cc}$.}
\label{fig_bandstructure}
\end{figure}

\subsection{Model}

To illustrate the resonant backscattering effect, we have performed
quantum transport calculations for two-terminal and six-terminal
nano-ribbon devices in a magnetic field. We ignore the effects of
inhomogenous doping and strain, which may also modify magnetotransport
as studied elsewhere, \cite{Prada:PRB10} and concentrate on the effect
of a bilayer patch. The starting point is the tight-binding
Hamiltonian for graphene
\begin{equation}
H  = \sum_{ij} t_{ij}c_i^{\dagger}c_j,
\label{eq_hamiltonian}
\end{equation}
where the hopping elements $t_{ij}$ include hopping beyond nearest
neighbors, and are modeled by the $\pi$-orbital overlap at different
carbon sites $j$ and $i$ separated by ${\bf R}_j-{\bf R}_i={\bf
  r}=(x,y,z)^T$,
\begin{equation}
t_{ij} = t({\bf r}) = -\gamma_0 \frac{x^2+y^2}{r^2}e^{-\lambda(|{\bf r}|-a_{cc})}
- \gamma_1 \frac{z^2}{r^2} e^{-\lambda(|{\bf r}|-d)}.
\label{eq_hopping}
\end{equation}
Here, $\gamma_0$ and $\gamma_1=0.14\gamma_0$ are the nearest neighbor
and interlayer hopping parameters of graphite, $a_{cc}$ is the
carbon-carbon distance in-plane and $d=2.4a_{cc}$ is the interlayer
distance. The exponent is $\lambda\approx 3/a_{cc}$.  The formula in
Eq.~(\ref{eq_hopping}) is applied for atomic distances $r=|{\bf r}|$
reaching a cut-off $R_c$, beyond which $t_{ij}=0$. This generalization
beyond simple nearest-neighbor models is crucial to properly recover
the low energy electronic structure of twisted bilayers, in particular
its gapless and valley-decoupled double-cone spectrum, as described by
the continuum theory of Ref. \onlinecite{Santos:PRL07}. In practice, a
rather precise description at relevant energy scales is obtained for
$R_c\gtrsim 7a_{cc}$.

\subsection{Bandstructure of the leads}

The band structure of the monolayer graphene nanoribbon leads
converges rapidly with increasing hopping cut-off $R_c$, and is shown
for a 10 nm wide zigzag nanoribbon with $R_c=3a_{cc}$ in
Fig.~\ref{fig_bandstructure} for varying magnetic fields.  The
magnetic field is included in the model through a standard Peierl's
substitution.  We note that for large $R_c$, a large unit cell of
length $a_u\geq R_c$ is needed for which the first Brillouin zone in
reciprocal space is small.  This correlates with the folding of the
bands of a nearest neighbor tight-binding model but leads to slightly
more complicated bands due to the long range hoppings, see
Fig.~\ref{fig_bandstructure}. For instance, for small magnetic fields,
Fig.~\ref{fig_bandstructure}(a), we see a positive energy shift of the
cones of about $0.3\gamma_0$ and the zero-energy edge modes of the
zigzag ribbon display substantial dispersion.\cite{Sasaki:1996} For
small magnetic fields $B$ the magnetic length
$\ell_B=\sqrt{\hbar/(|e|B)}$ is larger or comparable to the ribbon
width $W$ and the spectrum is dominated by size quantization. This is
the case in Fig.~\ref{fig_bandstructure}(a) where the energy split of
the zero-mode is due to the small magnetic field corresponding to a
flux $\Phi=10^{-3}\Phi_0$ per hexagon, where $\Phi_0=h/2e$ is the
magnetic flux quantum. For larger fields, the Landau levels
$E_n=\sqrt{2n}\hbar v_f/\ell_B = \sqrt{n}\omega_c$ ($v_f$ is the Dirac
electron velocity in the absence of magnetic field) become visible as
flat regions in the dispersion.  The dispersive parts of the bands
correspond to edge modes, carrying the current in the quantum Hall
regime.

\begin{figure}
\includegraphics[width=\columnwidth]{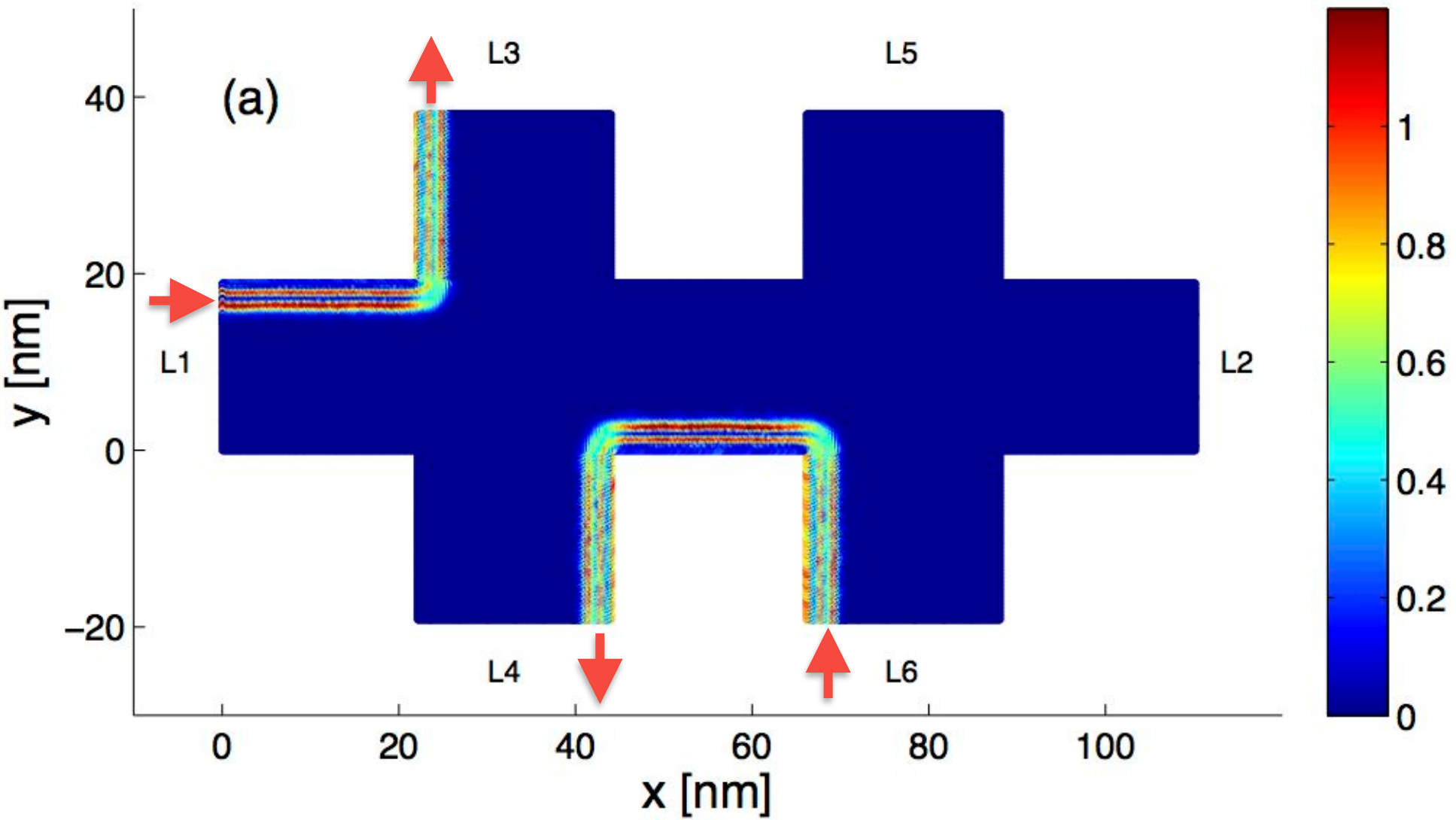}
\includegraphics[width=\columnwidth]{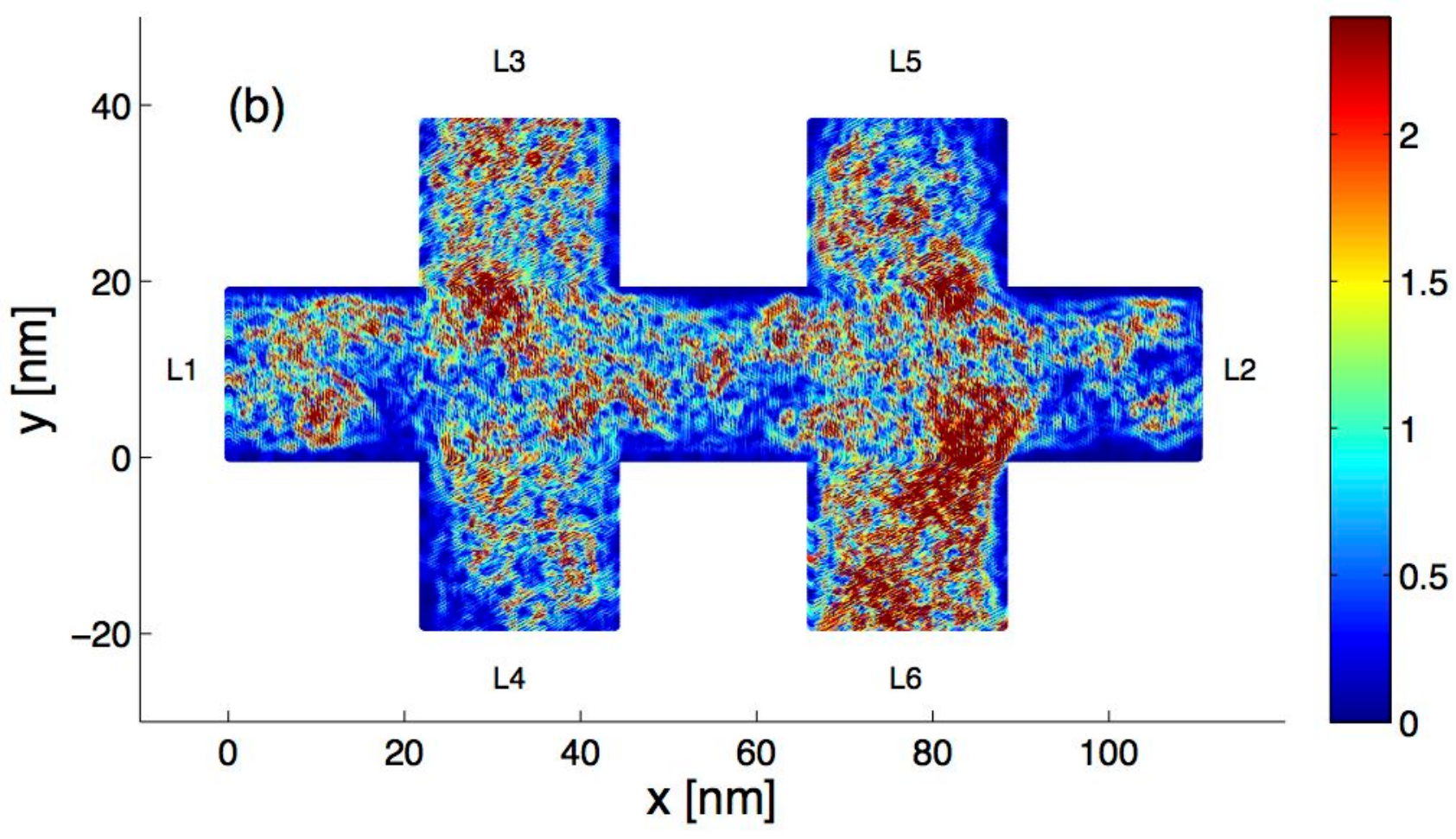}
\includegraphics[width=\columnwidth]{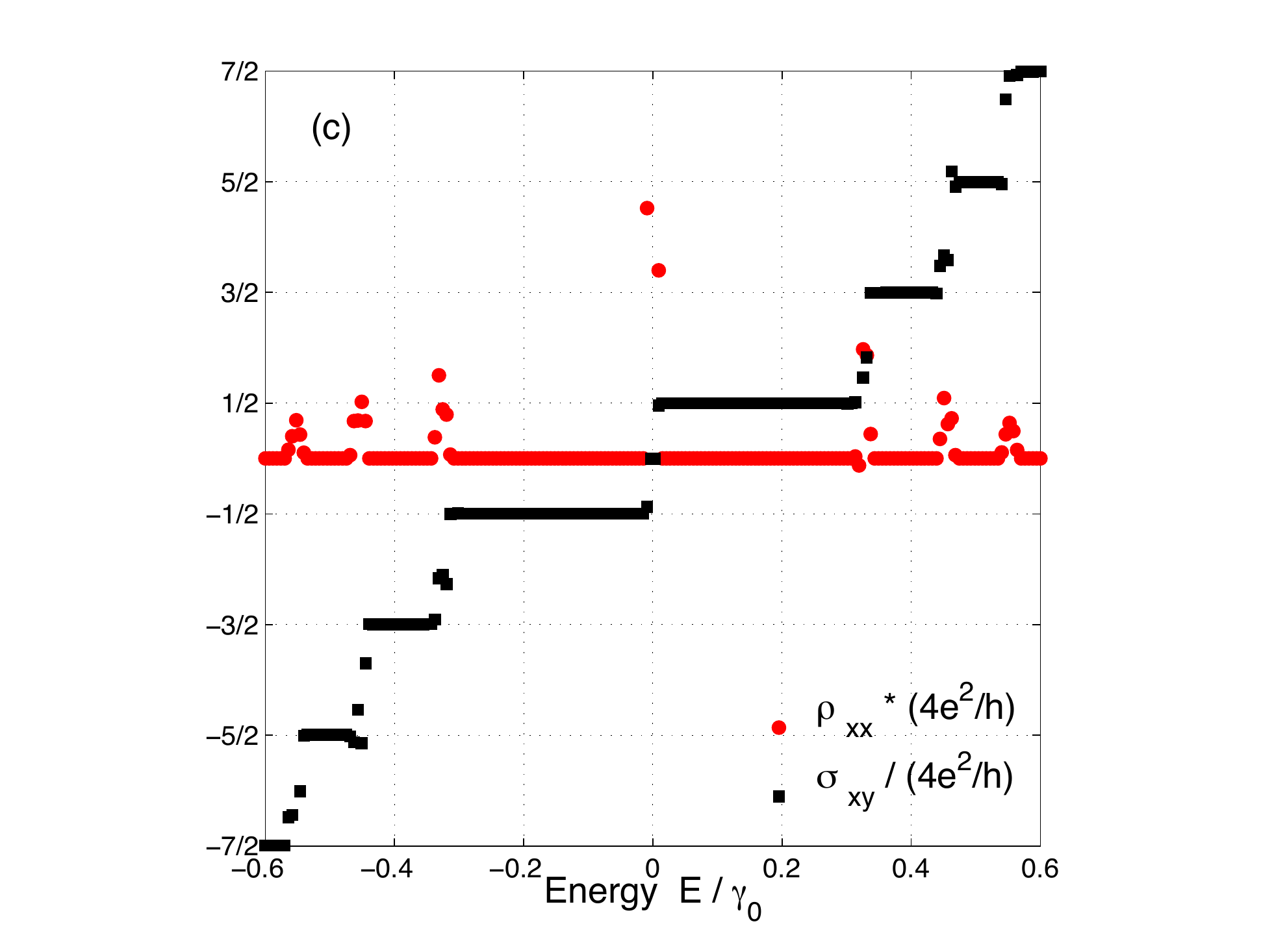}
\caption{(a) Carrier density in a monolayer graphene Hall bar with 6 leads, enumerated by
  L1-L6.  The Femi energy of the Hall bar is $E_F=0.5\gamma_0$, which
  corresponds to the $n=2$ Landau level [at the middle of the 3rd
  plateau in (c)]. Currents are injected at L6 and L1 and collected at
  L4 and L3. (b) The current flow patterns for $E_F=0.4507\gamma_0$,
  corresponding the step between plateaux $n=1$ and $n=2$.  (c) The
  longitudinal resistance $\rho_{xx}$ (red dots; voltage measured
  between L3 and L5) and the transverse conductance $\sigma_{xy}$
  (black squares; voltage measured between L5 and L6).  The color
  scale in (a) and (b) are given in units of $G_0V$, where V is the
  small increase of the chemical potentials in L6 and L1 with respect
  to the other leads. The applied field corresponds to a flux
  $\Phi=0.01\Phi_0$ per hexagon, and the temperature is zero.}
\label{fig_Hall}
\end{figure}

\section{Magnetotransport simulation}

To evaluate the effect of the bilayer patch, we compute
magnetotransport properties using recursive Green's function (RGF)
techniques, \cite{Datta:97,Waintal:2008} where the coupling to
reservoirs are included through self-energies derived from the surface
Green's functions of semi-infinite leads.  The leads and the system
are modeled on equal footing through the Hamiltonian in
Eq.~(\ref{eq_hamiltonian}). The RGF algorithm gives the retarded
Green's function of the system.  Such Green's function is obtained,
between certain pairs of points, by iterative application of the Dyson
equation, and may be then used to compute the Hall conductivity, the
current densities or the scattering matrix of the system. The
recursive iteration is performed on slices of the lattice that are
connected only to neighboring slices, and which hence increase in size
as the hopping cut-off $R_c$ increases. This has a rather steep
computational cost, but has the advantage that it cleanly avoids
fermion doubling problems that plague strategies based on the
discretization of low energy effective theories in graphene, and can
moreover quantitatively incorporate the precise edge termination of
each of the Hall bar regions.

\subsection{Multiterminal Hall conductivity}
\label{section_six-terminal}

In Fig.~\ref{fig_Hall} we display a 6-terminal monolayer graphene Hall
bar device with six contacts (leads) enumerated by L1-L6.  In a
typical experiment, a current is sent from L1 to L2, and the voltage
between L3 and L5 gives the longitudinal resistance, while the Hall
resistance is obtained by measuring the transverse voltage between for
instance L5 and L6. Since this Hall bar is of monolayer graphene only,
it is sufficient to use a nearest neighbor model.  After computation
of the full scattering matrix connecting the six leads, we compute the
longitudinal resistance $\rho_{xx}$ and the transverse conductance
$\sigma_{xy}$ in the linear response regime.  We display both in (c)
as function of Fermi energy of the system (related to the electron
density). The transverse conductance display quantized values
$\sigma_{xy}=\pm(2n+1)\,G_0$, where {$G_0=2e^2/h$ and $n=0,1,2,...$
  This sequence is characteristic for the monolayer quantum Hall
  effect.  The longitudinal resistance is zero except at the steps
  between plateaux. The random fluctuations at the steps are due to
  the added randomness of 10\% of the nearest neighbor hopping
  integral $t_{ij}$ around $\gamma_0$ in this simulation.

  The current flows along edge states, as is clearly seen in
  Fig.~\ref{fig_Hall}(a), which shows the local current flow patterns
  throughout the device when currents are injected at L6 and L1 and
  subsequently collected at L4 and L3.  In Fig.~\ref{fig_Hall}(b) we
  show the current redistribution throughout the entire device that
  appears at each step between plateaux (in this case the $n=1$ and
  $n=2$ plateaux at $E_F=0.4507\gamma_0$).

  In Fig.~\ref{fig_Hallpatch} we show the influence of an AB-stacked
  bilayer stripe defect placed in the middle and connecting the two
  edges of the Hall bar. The current can now enter into a circular
  path around the bilayer patch and eventually go out into both leads
  L5 and L4, see Fig.~\ref{fig_Hallpatch}(a).  This leads to large
  fluctuations of the longitudinal resistance $\rho_{xx}$, as shown in
  Fig.~\ref{fig_Hallpatch}(a), red circles. At the same time, the
  transverse conductance is affected. If the voltage probes are set
  between L5 and L6, the influence of the patch is minimal.  On the
  other hand, when the voltage probes span the patch, for instance
  when they are placed between L3 and L6, the fluctuations are added
  into $\sigma_{xy}$ as well, and the plateaux are destroyed.

\begin{figure}
\includegraphics[width=\columnwidth]{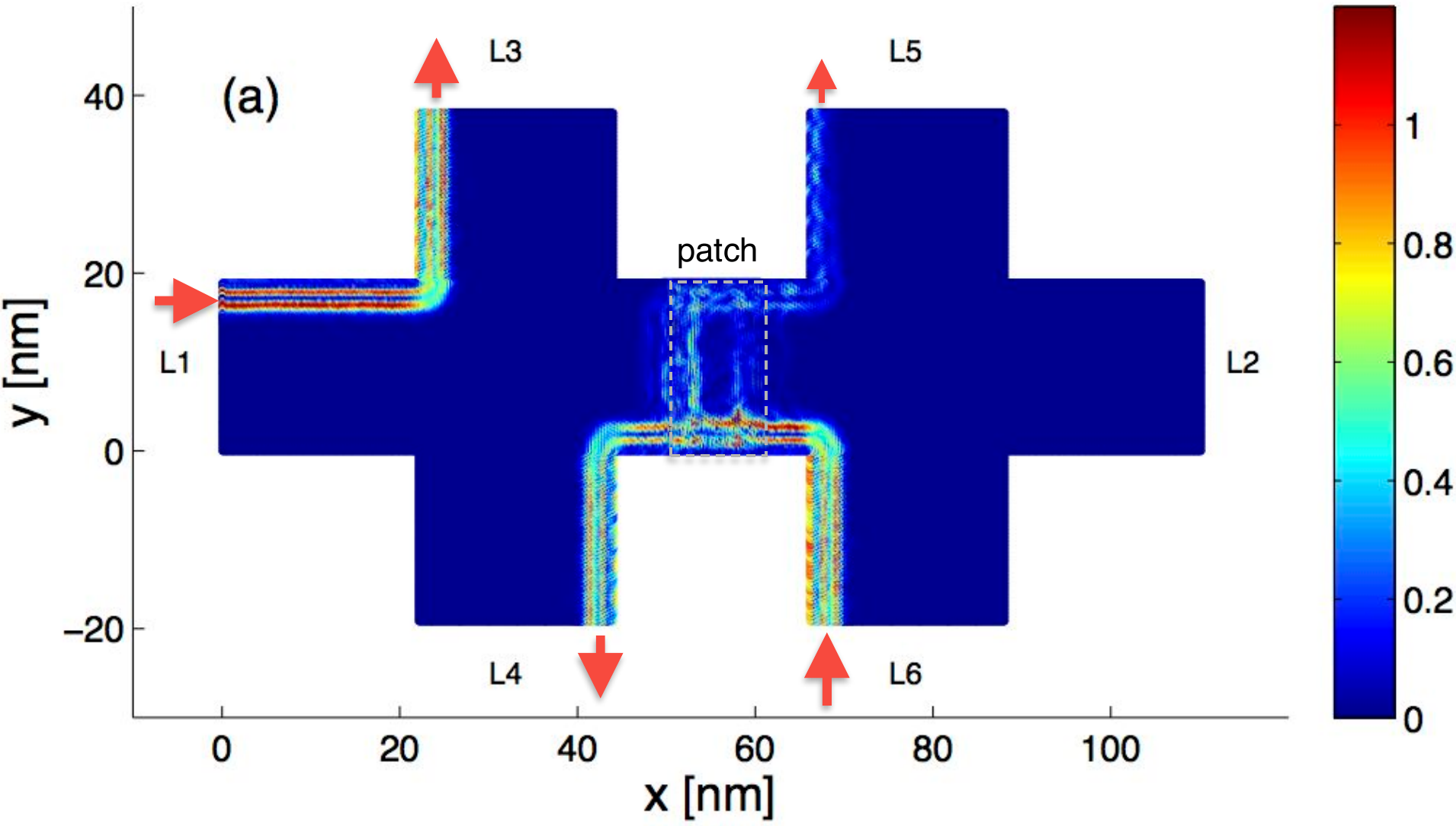}
\includegraphics[width=\columnwidth]{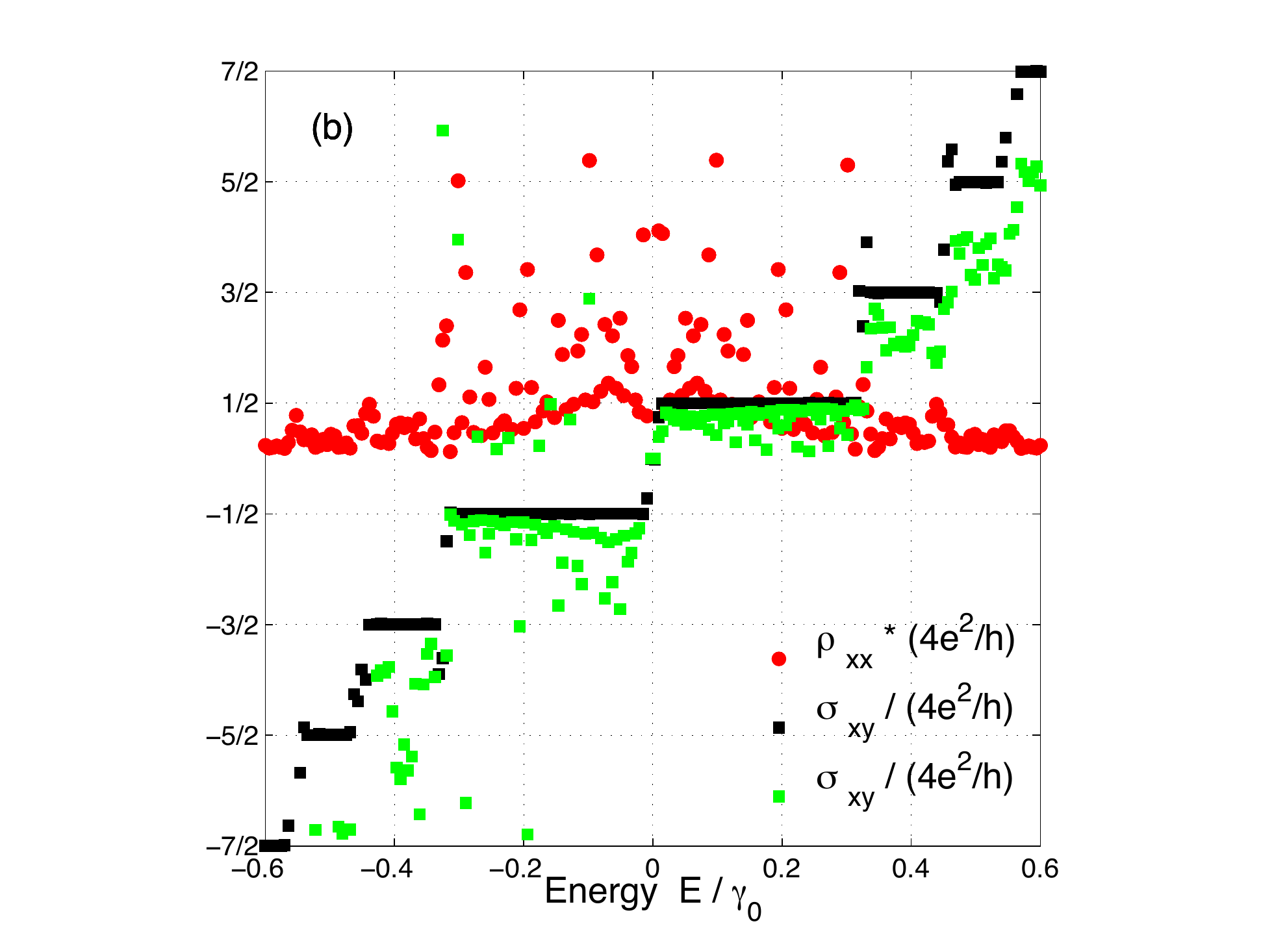}
\caption{(a) A Hall bar with a $L=10$ nm long AB-stacked bilayer
  stripe defect in the middle of the device, connecting the two edges
  at $y=0$ and $y=20$ nm.  The local current flow pattern is for the
  Femi energy $E_F=0.5\gamma_0$ [the same as in
  Fig.~\ref{fig_Hall}(a)].  (b) The longitudinal resistance
  $\rho_{xx}$ (red dots; voltage measured between L3 and L5), and the
  transverse conductance $\sigma_{xy}$ computed for a voltage measured
  either between L5 and L6 (black squares) or between L3 and L6 (green
  squares).  The color scale in (a) is given in units of $G_0V$, where
  V is the small increase of the chemical potentials in L6 and L1 with
  respect to the other leads.  The applied field corresponds to a flux
  $\Phi=0.01\Phi_0$ per hexagon, and the temperature is zero.}
\label{fig_Hallpatch}
\end{figure}

\subsection{Two-terminal conductance and twist angle}

Next we study the influence of a finite twist angle on the
fluctuations of the Hall conductance.  In contrast to the AB-stacked
bilayer patch explored above, this requires that we include long-range
hopping $t_{ij}$ with a cut-off $R_c=7a_{cc}$. For this study it is
convenient to limit the calculations to a two-terminal set-up, as in
Fig.~\ref{fig_system}(b).  In the absence of contact resistances, like
in the present case, the two-terminal conductance equals the Hall
conductivity $\sigma_{xy}$.  In Fig.~\ref{fig_conductance} we show the
conductance for a $W=10$ nm wide ribbon in a quantizing magnetic field
as function of Fermi energy (i.e. doping).  The bandstructure of the
underlaying monolayer is shown in Fig.~\ref{fig_bandstructure}(d).
The bilayer patch length is $L=10$ nm.  The twist angle $\theta=0$
corresponds to a bilayer patch with AB stacking, while
$\theta=60^\circ$ would correspond to AA stacking.  For small twist
angles, all plateaux are destroyed by backscattering caused by a
number of resonance states in the patch, which are rather broad and
tend to overlap.  For increasing twist angle, these resonances become
sharper, signaling a decoupling of the quasibound states from the edge
modes in the underlying monolayer that is connected to source and
drain reservoirs.  The plateaux become better defined, starting with
the $n=0$ plateau at small twist angle and continuing with the higher
Landau level index plateaux at higher twist angles (higher index
require larger twist angle to recover). \footnote{Note that the peak
  in Fig.~\ref{fig_conductance} around $0.25\gamma_0$ (in the zeroth
  plateau), is related to the dispersive zero mode characteristic of
  the zig-zag nanoribbon of Fig. \ref{fig_bandstructure}(d).}

The dependence with twist angle of the width of the backscattering
resonances, or in other words, of the coupling between the
corresponding quasibound state and the monolayer edge states, can be
traced to the band structure of the twisted bilayer.  In the limit of
a vanishing interlayer coupling $\gamma_1\ll\gamma_0$, the states
become perfectly bound and lie fully on the decoupled layer.  Their
momentum components are concentrated around the Dirac point of said
layer.  In contrast, the delocalized edge states in the extended
monolayer are spectrally concentrated around the monolayer Dirac
point, which has a shift $\Delta
K=2\sin(\theta/2)\times4\pi/(3\sqrt{3}a_{cc})$ [for
$0<\theta<30^\circ$] with respect to the former.  The momentum spread
grows linearly with energy. Therefore, for a given energy, the larger
the momentum mismatch of the two Dirac points, the smaller the overlap
between delocalized edge states and localized patch states will
be. Since this overlap is a measure of the inverse lifetime of the
quasibound state in the limit of small $\gamma_1/\gamma_0$, we see
that twist angles around $30^\circ$ (half-way between AB and AA
stacking, maximum $\Delta K$) will correspond to least coupling,
narrower resonances, and cleaner Hall plateaux, as seen in
Fig.~\ref{fig_conductance}.

A second consequence of this analysis is that, as soon as the Fermi
energy approaches the van Hove singularity where the two Dirac cones
intersect (at energy $\sim v_F\Delta K/2 - \gamma_1$), the spectral
spread becomes comparable to $\Delta K$, so the overlap will increase
greatly, and the backscattering will be enhanced. Hence, higher Hall
plateaux will be eventually destroyed for any value of the twist angle
as the filling factor grows. This is also apparent in
Fig.~\ref{fig_conductance}(a). For instance, for $\theta=20^\circ$
(blue curve), it is clear that the $n=0$ and $n=1$ plateaux have sharp
resonances, while plateau $n=2$ and especially plateaux $n\geq 3$ at
higher doping (i.e. higher $E/\gamma_0$) are destroyed.

\begin{figure}[t]
\includegraphics[width=\columnwidth]{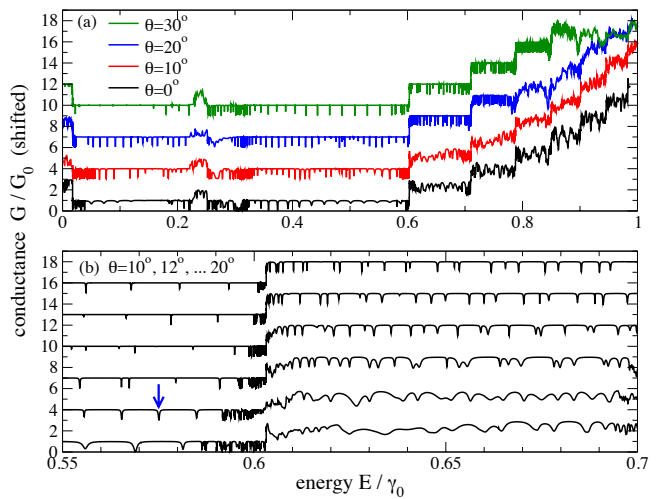}
\caption{(a) Conductance at zero temperature for a 10 nm wide zigzag
  ribbon with a 10 nm long bilayer patch at various twist angles
  $\theta$.  The magnetic field corresponds to a flux
  $\Phi=0.01\Phi_0$ per hexagon.  The hopping cut-off is
  $R_c=7a_{cc}$.  (b) Energy blow-up of (a) around the first
  conductance step from Landau level $n=0$ to $n=1$.  The blue arrow
  in the figure points to the conductance fluctuation at which we
  display the local current flow patterns in Fig.~\ref{fig_local}.
  The curves are shifted by $3G_0$ relative to each other for
  clarity.}
\label{fig_conductance}
\end{figure}

\subsection{Circulating quasi-bound states}

To demonstrate the connection between resonant backscattering and
quasibound states of circulating currents around the patch, we present
in Fig.~\ref{fig_local} the local current flow pattern throughout the
system for filling factors near and at the resonance dip indicated by
the blue arrow in Fig.~\ref{fig_conductance}(b). In the first panel
the edge current flows from left to right, from source to drain, along
the upper edge in the $n=0$ Landau level of the monolayer undisturbed
by the patch. On resonance (fourth panel) the current circulates in
the patch and suffers perfect back reflection at the lower edge (blue
back-flowing current), and the conductance from such edge state is
zero on resonance.  Similar resonances occur at higher plateaux, where
resonant backscattering from each patch state always removes at most
one conductance quantum $G_0$ from the Hall conductivity (assuming
unbroken spin symmetry).

\begin{figure}[t]
\includegraphics[width=\columnwidth]{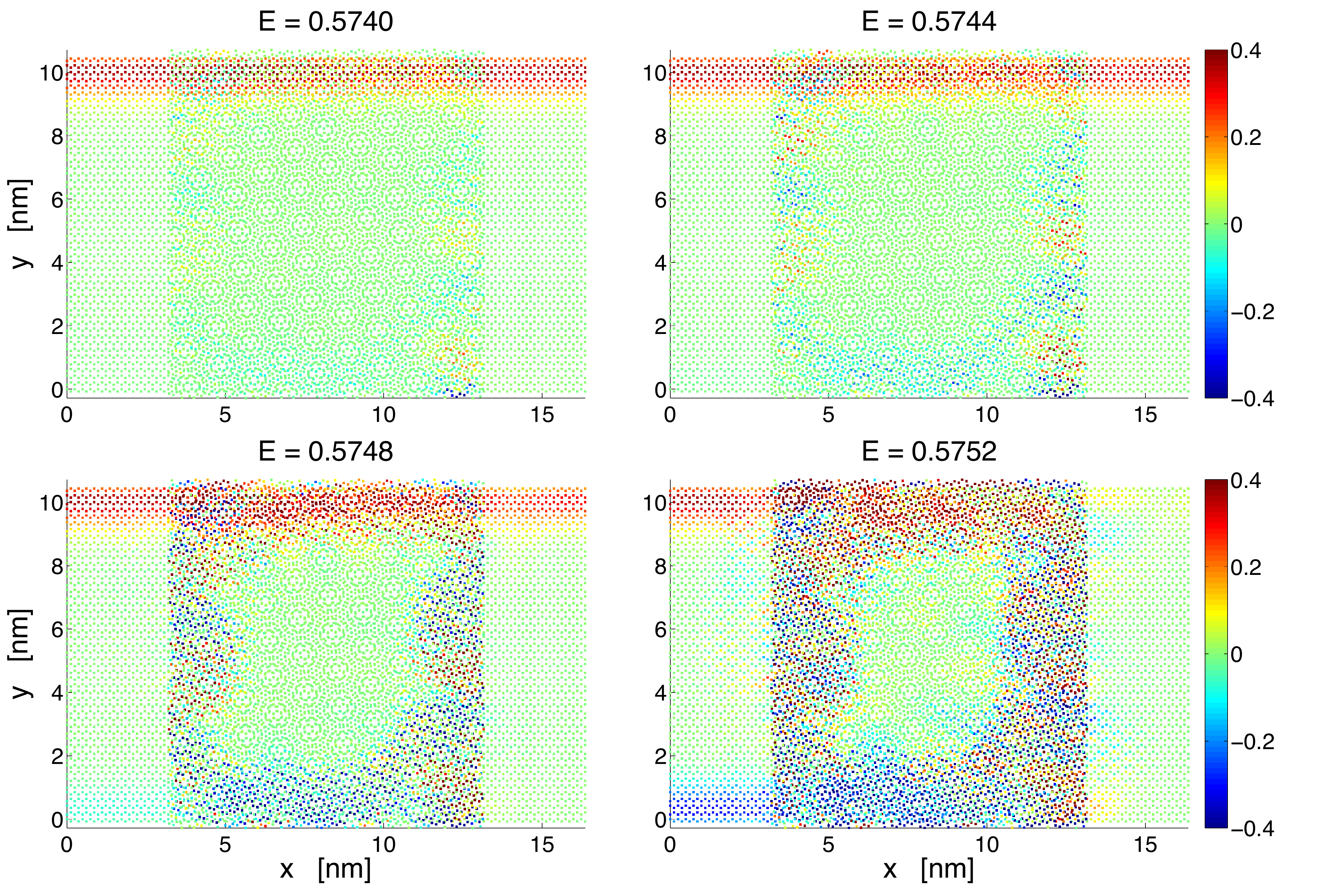}
\caption{Local current flow (x-component of the bond currents in units
  of $G_0V$) at zero temperature at a twist angle $\theta=12^\circ$.
  The four frames corresponds to energies near the conductance
  fluctuation indicated by the blue arrow in
  Fig.~\ref{fig_conductance}(b), starting at an energy below the
  conductance dip on the plateau and ending near the minimum of the
  dip.  Model parameters are the same as in
  Fig.~\ref{fig_conductance}. The source and drain reservoirs are
  located to the left and right of the device and the current flows
  from left to right for red color (positive sign).}
\label{fig_local}
\end{figure}

\subsection{Effects of Disorder}

In the above two-termainal simulations of the bilayer stripe defect we
have neglected disorder. A simple model of disorder was included in
the 6-terminal simulations in Section~\ref{section_six-terminal}
through a $10\%$ randomization of the nearest neighbor hopping
integrals $t_{ij}$ around $\gamma_0$ in that case. To simulate the
influence of disorder on the conductance
fluctuations induced by the bilayer patch, we do the same for the $t_{ij}$ in
Eq.~(\ref{eq_hopping}).  In practice, we make the substitution
$t_{ij}\rightarrow t_{ij}(1+\lambda\rho_{ij})$, where the level of
disorder is $\lambda$ and $\rho_{ij}$ is a random number between
$-0.5$ and $0.5$. In Fig.~\ref{fig_disorder} we show results of this
type of disorder for a bilayer patch with twist angle
$\theta=20^\circ$ for energies $E\in[0.63,0.66]\gamma_0$, which is on
the $n=1$ plateau in Fig.~\ref{fig_conductance}. We vary the disorder
strength from $\lambda=5\%$ to $\lambda=30\%$, with the same random
number sequence $\rho_{ij}$. For small disorder strength, the
resonances are shifted in energy. For increasing disorder strength
(bigger $\lambda$), resonances get broadened. Eventually, resonances
overlap and the plateau is completely destroyed again, despite its
$\theta=20^\circ$ twist angle. This destruction can be understood as
due to enhanced momentum relaxation that reduces the effect of the
momentum mismatch between the Dirac cones of the two layers that in
the first place (without disorder) decoupled the layers and lead to
sharp resonances. We find, thus, that a sizeable amount of disorder is
necessary to cause an appreciable correction to the general results
found in the clean case.

\begin{figure}[t]
\includegraphics[width=\columnwidth]{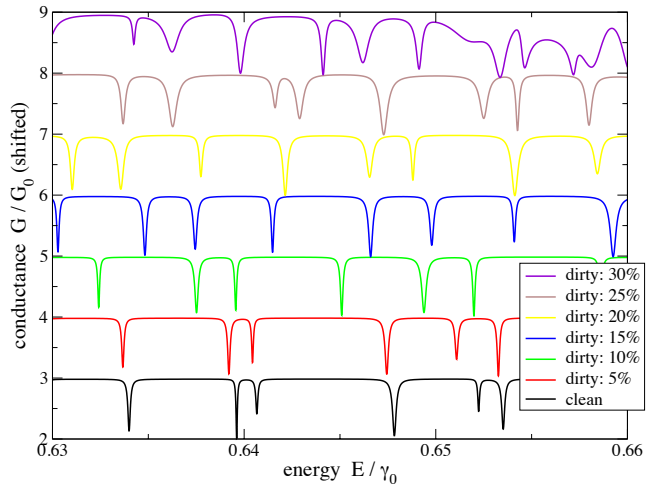}
\caption{The effect of disorder on the conductance fluctuations for a
  bilayer patch with twist angle $\theta=20^\circ$. With increasing
  amount of disorder, conductance dips are shifted and
  broadened. Eventually, dips overlap and the plateau is destroyed
  despite the large twist angle.}
\label{fig_disorder}
\end{figure}

\section{Conclusions and Outlook}

We have analyzed the effect of bilayer stripes transverse to graphene
Hall bars on the Hall conductivity. Such stripes are observed to
naturally arise at substrate steps in epitaxially grown graphene. We
have found that, in agreement with Ref. \onlinecite{Schumann:PRB12},
the Hall plateaux are destroyed by the coupling between opposite edge
states via transverse transport channels circulating around the
bilayer perimeter.  Such channels arise as a result of the jump in
Chern number between the band structures of bilayer and monolayer
graphene, and give rise to the formation of circulating quasibound
states in finite length bilayer patches. Hall plateaux develop
backscattering resonances, visible as dips of depth one conductance
quantum, whenever the Fermi energy crosses a quasi bound level in the
patch. The width of the backscattering resonances diminishes as the
bilayer twist angle approaches $30^\circ$, which leads to well defined
low energy plateaux despite the patch. However, resonance width grows
with Fermi energy, completely spoiling Hall plateaux above the van
Hove singularity of the twisted bilayer patch.  Both features are
explained in terms of the momentum mismatch between the Dirac cones in
the two layers. We propose that this scattering mechanism should be
relevant in understanding deviations of the QHE in epitaxial graphene
Hall bars etched across SiC steps, like the anomalous positive
magnetoresistance and non-quantized Hall plateaux in
Ref. \onlinecite{Schumann:PRB12}.

In this first study of such Hall bars, sketched in
Fig.~\ref{fig_system}(a), we have neglected effects of strain,
inhomogeneous doping, and the possibility of a Zeeman term due to an
in-plane component of the magnetic field. In addition, we have
neglected electron-electron interactions that may lead to wider
wavefunctions of the edge states [the current paths in for instance
Fig.~\ref{fig_Hall}(a)]. The relavance of these effects, which should
be present at least to some extent in real experiments, are left as
future work. However, our expectation is that, since the essential
mechanism for the low energy protection of Hall plateux found in our
work stems from the momentum mismatch between layers, the destructive
effect of inhomogeneities, including those arising from strain and
screening, will be small, as long as their characteristic lengthscales
are greater than the Moir\'e period of the twisted bilayer patch
$L_M=\sqrt{3}a_{cc}/2\sin(\theta/2)$.

\acknowledgments We acknowledge financial support from the EU through
FP7 STREP ConceptGraphene (T.L.), the Swedish Foundation for Strategic
Research (T.L.), the CSIC JAE-Doc program and the Spanish Ministry of
Science and Innovation through Grant No.  {FIS2011-23713} (P.S.-J) and
FIS2009-08744 (E.P.), and the European Research Council Advanced
Grant, contract 290846 (P. S-J.).  This research was supported in part
by the National Science Foundation under Grant No. NSF PHY05-51164.

\end{document}